\documentclass{iopart}
\usepackage{iopams}
\usepackage[dvips]{graphicx}% Include figure files
\usepackage{bm}% bold math
\usepackage{amsfonts}
\usepackage{amsgen}
\usepackage{amsbsy}
\usepackage{hyperref}% makes the references hyperlinked

\begin{document}
\title[Energy and Hamiltonian]
{Energy and Hamiltonian in first order gravity}
\author{Rodrigo Aros}
\address{Universidad Andr\'es Bello, Republica 252, Santiago,Chile} \ead{raros@unab.cl}
\begin{abstract}
In this work the definition of a quasilocal energy for four dimensional first order gravity is
developed. Using this an action principle which is adequate for the canonical ensemble is
obtained. The microcanonical action principle is obtained as well.
\end{abstract}
 \submitto{\CQG}

%04.60Quantum gravity, 04.60.Gw Covariant and sum-over-histories quantization

\section{Introduction}

The role of boundary conditions and their connection with the boundary terms has been addressed in
many works. Essentially the boundary terms must be such that the variation of the action, provided
the boundary conditions, vanish in order to have a well defined variational principle. Furthermore
those boundary conditions also must be such that the equations of motion might have solutions.

However there is a third role which arises from certain formal relations that connect quantum
field theory with statical mechanics. For instance the partition function in the canonical
ensemble can be written as
\begin{equation}
Z(\beta) =  \oint \mathcal{D} x e^{i I|_{\tau=-i\beta}},
\end{equation}
provided $I$ effectively be such that the temperature $\beta^{-1}$, said the inverse of the
period, be fixed. In principle there is a suitable action for the microcanonical ensemble or for
any other ensemble.

For gravity, although there is no known quantum gravity, analogous results at tree level
remarkably seem to work \cite{Hawking:1983dh}. Those somehow mysterious results justify to pursue
a Hamiltonian analysis for gravity. For gravity this has been addressed in many different ways,
for instance \cite{Regge:1974zd,Brown:1993bq,Peldan:1994hi}.

\subsection{First Order gravity}
However the metric formulation of gravity lacks of a very important characteristic, fermions can
not be incorporated because, roughly speaking, the group of diffeomorphism does not have half
integer representations. To incorporate those fields is necessary to introduce a local orthonormal
basis, namely a vierbein which is usually written in terms of the set of differential forms
$e^{a}=e^{a}_{\mu} dx^{\mu}$, because implicitly one is incorporating a local Lorentz group where
fermions can be realized.

The introduction of the vierbein leads to a different formulation of gravity
\cite{Zumino:1986dp,Regge:1986nz}. In this formulation the corresponding Lorentz connection, which
is also written in terms of the differential forms $\omega^{ab}=\omega^{ab}_{\mu}dx^{\mu}$ called
the spin connection, is an additional independent field. This formulation is essentially different
in many aspects, for instance in four dimensions although the number of degrees of freedom in both
formulation is 2 the standard 10 fields contained in the metric in this new version of gravity are
replaced by 40 fields, or besides the usual constraints of diffeomorphisms another that fixes the
Lorentz rotations arises. This new formulation is usually called the Einstein Cartan Gravity or
because of its first order equations of motion is called first order gravity as well.

The four dimensional Einstein Hilbert action in first order formalism reads
\begin{equation}\label{EHaction}
    I_{EH} = \int_{\mathcal{M}} R^{ab}\wedge e^{c}\wedge e^{d}\varepsilon_{abcd},
\end{equation}
where \[R^{ab}=d\omega^{ab} + \omega^{a}_{\hspace{1ex} c} \wedge \omega^{cb}=
\frac{1}{2}R^{ab}_{cd}e^{c}\wedge e^{d},\] being $R^{ab}_{cd}$ the Riemann tensor.
$\varepsilon_{abcd}=\pm 1,0$ stands for the complete antisymmetric symbol, which differs form
Levi-Civita pseudo tensor by $\sqrt{g^{(4)}}$. To avoid confusion in this work only antisymmetric
symbols will be used and any determinant of the metrics will be written explicitly.

The variation of Eq.(\ref{EHaction}) yields the two set of equations of motion
\begin{eqnarray}
  \delta e^{d} &\rightarrow& R^{ab}\wedge e^{c}\varepsilon_{abcd}=0 \label{EinsteinEquations}\\
  \delta \omega^{ab} &\rightarrow& T^{c}\wedge e^{d}\varepsilon_{abcd}=0\label{TorsionEquations},
\end{eqnarray}
where $T^{a}=De^{a}=\frac{1}{2}T^{a}_{bc}e^{b}\wedge e^{c}$ corresponds to the torsion two form
and $T^{a}_{bc}$ is the torsion tensor. Note that Eq.(\ref{TorsionEquations}) is an algebraic
equation, thus its solution is $T^{a}=0$. This combined with that Eq.(\ref{EinsteinEquations}) are
indeed the Einstein equations implies that the solutions of the metric formalism are recovered
on-shell by this formulation.

Note from Eq.(\ref{EinsteinEquations}) that in this case the \textit{energy momentum} tensor has a
mixed nature, having indexes in the orthonormal and coordinate bases, namely
\begin{equation}\label{EMtensor}
    \frac{\delta I}{\delta e^{a}_{\mu}} =  T^{\mu}_{a}.
\end{equation}
One important difference of this energy momentum tensor (\ref{EMtensor}) is that even if it is
transformed into a single basis -either the coordinate basis or the orthonormal bases- this tensor
is not symmetric, as in the metric formalism, unless the $T^{a}=0$ condition be imposed.

Four dimensional first order gravity differs from any higher dimensional first order gravity in
very important aspect. In general $e^{a}$ and $\omega^{ab}$ are independent fields, and thus one
can expect that there is an independent conjugate momentum for each one. However in four
dimensions the conjugate of momentum of $e^{a}$ is contained in $\omega^{ab}$ or viceversa.
\cite{Banados:1997hs,Contreras:1999je}. This leads to the definitions of two equivalent phase
spaces that can be mapped into each other readily where to develop the Hamiltonian theory. They
are called $e$ and $\omega$-frames respectively.  Furthermore the equivalence of their path order
integrals have been confirmed \cite{Contreras:1999je}. In terms of this notation in this work the
e-frame will be used exclusively.

\subsection{Energy}

The quest for a definition of energy for gravity has been addressed by many authors in many forms
\cite{Komar1959,Regge:1974zd,Lee:1990nz,Wald:1997qz,Henneaux:1999ct,Aros:1999id,Barnich:2001jy,Barnich:2003xg}.
This even for classical mechanics relays on the boundary terms of the action. Those terms fix the
ground state of the system, and thus they fix zero of energy, but also a definition of a finite
energy might relay on them as well (see for instance \cite{Hawking:1996fd,Aros:1999id}). In this
work the definitions of a quasilocal energy for four dimensional first order gravity will be
studied trying to shed some more light into the problem of energy in gravity.

It is worth to mention another approach to this subject in \cite{Fatibene:2000jm}, where another
kind of first order gravity is discussed.

\subsection{The space}

The space to be discussed in this work corresponds to a topological cylinder. One can picture it
as $\mathcal{M}=\mathbb{R}\times \Sigma$ where $\Sigma$ corresponds to a 3-dimensional spacelike
hypersurface and $\mathbb{R}$ stands for the time direction and formally is a segment of the real
line. In this way the boundary of the space is given by $\partial\mathcal{M}=
\mathbb{R}\times\partial\Sigma\cup \Sigma_{+} \cup \Sigma_{-}$, where $\Sigma_{\pm}$ are the upper
and lower boundaries of the topological cylinder and $\partial \Sigma$ is the boundary of
$\Sigma$.

\section{Fixing the fields}
The variation of the EH action (\ref{EHaction}) yields the boundary term
\begin{equation}\label{Variation}
\delta I_{EH}|_{\textrm{on shell}} =  \int_{\mathcal{\partial M}} \delta \omega^{ab}\wedge
e^{c}\wedge e^{d}\varepsilon_{abcd},
\end{equation}
which implies that the pure EH action would be a proper action principle provided
$\delta\omega^{ab}$ vanish at the boundary. This -on shell- is equivalent to fix the extrinsic
curvature at the boundary, which for reasons that would clear later is not a convenient boundary
condition in this case. In the metric formalism to achieve the quasilocal energy definitions, and
finally the canonical ensemble action, the corresponding boundary condition would fix the metric
at $\partial \mathcal{M}$ \cite{Brown:1992br}. With this in mind here it will be choose as
boundary condition the fixing of the vierbein. This argument leads to modify the action as
\begin{equation}\label{EHactionimproved}
I_{EH}\rightarrow \tilde{I}_{EH} = \int_{\mathcal{M}} R^{ab}\wedge e^{c} \wedge
e^{d}\varepsilon_{abcd}- \int_{\mathcal{\partial M}}  \omega^{ab}\wedge e^{c}\wedge
e^{d}\varepsilon_{abcd},
\end{equation}
which yields
\begin{equation}\label{VariationImproved} \delta
\tilde{I}_{EH}|_{\textrm{on shell}} = - 2 \int_{\mathcal{\partial M}} \omega^{ab}\wedge
e^{c}\wedge \delta e^{d}\varepsilon_{abcd},
\end{equation}
justifying that the action (\ref{EHactionimproved}) is a proper action principle provided $e^{a}$
is fixed at the boundary.

It is worth to mention that the term added to the action -on shell- is actually the standard
extrinsic curvature term, \textit{i.e.},
\begin{equation}\label{ExtrisicTerm}
\int_{\mathcal{\partial M}}  \omega^{ab}\wedge e^{c}\wedge e^{d}\varepsilon_{abcd} =
\int_{\mathcal{\partial M}} K \sqrt{h}\, d^{3}y
\end{equation}
where $K$ is the trace of the extrinsic curvature of either $\Sigma_{\pm}$ or $\mathbb{R}\times
\partial \Sigma$ respectively, $h$ the determinant of the induced metric and $y$ an adequate coordinate system.

\section{First order gravity in Hamiltonian}

To achieve a Hamiltonian prescription of the first order gravity its necessary to define adequate
vielbein and coordinate systems. To begin, one can follow the standard method and use the line
element \cite{Arnowitt:1962hi}
\begin{equation}\label{ADM}
    ds^{2} = -N^{2} dt^{2} + g_{ij}(N^{i} dt + dx^{i})(N^{j} dt + dx^{j}).
\end{equation}
The inverse of the metric in Eq.(\ref{ADM}) is
\begin{equation}\label{ADMinverse}
    g^{\mu\nu} = \left[\begin{array}{ll}
             -N^{-2} &  N^{i} N^{-2} \\
                         N^{j} N^{-2} & g^{ij}_{(3)}-N^{i}N^{j}N^{-2}
                       \end{array}\right]
\end{equation}
where $g^{ij}_{(3)}g_{jk} = \delta^{i}_{k}$.

Next one splits the coordinates $x^{\mu}$ into time and space, i.e. $x^{\mu} = (t,x^{i})$, which
allows to rewrite (\ref{EHactionimproved}) as
\begin{equation}\label{EHactionimprovedsplit}
\tilde{I}_{EH} = \int_{\mathcal{M}} (\dot e^{a}_{i} \, \Omega^{\, i\  j}_{a\ bc} \,
 \omega^{bc}_{j}+\omega^{a b}_{0} J_{a b} + e^{a}_{0} P_{a})\,dt\wedge d^3x + B
\end{equation}
where
\begin{eqnarray*}
B&=&\int_{\mathbb{R} \times \partial \Sigma} 2\,e^{a}_{t}\, \Omega_{a\ bc}^{\, i\
j}\,\omega^{bc}_{j} \varepsilon_{imn}\,dt\wedge dx^{m}\wedge dx^{n}\\
  J_{ab} &=& 2\, T^{c}_{ij} e^{d}_{k} \varepsilon_{abcd}\varepsilon^{ijk} \\
  P_{d}&=& 2\, R^{ab}_{ij} e^{c}_{k}\varepsilon_{abcd}\varepsilon^{ijk},\\
  \Omega_{a\ bc}^{\, i\  j} &=& 2 \varepsilon_{abcd} \varepsilon^{ijk} e^d_k.\label{Omega}
\end{eqnarray*}
Note that the action (\ref{EHactionimprovedsplit}) has only a boundary term at $\mathbb{R}\times
\partial \Sigma$.

Recalling that the vielbein is fixed at the boundary, \textit{i.e.}, $\delta
e^{a}_{\mu}|_{\partial \mathcal{M}}=0$, the variation of the action with respect to $e^a_t$ and
$w^{ab}_t$ gives the constraint equations,
\[
P_a=0   \textrm{   and   }     J_{ab}=0.
\]

Now, it is a matter of fact that $J_{ab}$ can be interpreted as the generator of Lorentz
transformations and $P_{a}$ as the generator of translations \cite{Banados:1997hs}.

To continue one needs to define the vierbein. Among the different vierbeinen that give rise to
Eq.(\ref{ADM}) here it will be used \cite{Peldan:1994hi}
\begin{eqnarray}
  e^{a}_{t} &=& N \eta^{a} + N^{i} e^{a}_{i} \nonumber\\
  e^{a}_{i} &=& e^a_{i}, \label{ADMlabel}
\end{eqnarray}
with
\[
\eta^{a} e_{ai} = 0,\textrm{       }e_{ai}e^{a}_{j}= g_{ij},\textrm{ and    }\eta^{a}\eta_{a}=-1,
\]
because it significatively simplifies the computations. $\eta_{a}$ is the unitarian vector normal
to the $t=cont.$ slices $\Sigma$. In four dimensions $\eta^{a}$ can be constructed as
 \[\eta_{a} = \frac{1}{6\sqrt{g}}\,
\varepsilon_{abcd}\,e^{b}_{i}e^{c}_{j}e^{d}_{k}\varepsilon^{ijk},\] where $g = \det{g_{ij}}$. The
construction of $\eta^{a}$ in higher dimensions is straightforward.

Before to continue one it is worth to note that the transformation (\ref{ADMlabel}) is in fact the
transformation of fields
\[
(e^{a}_{t},e^{a}_{i}) \rightarrow (N,N^{i},e^{a}_{i}),
\]
which can be proven to be of unitarian determinant\cite{Aros:2003bi} in the space of
configurations, namely this transformation does not introduce a change in the measure in the path
order integral.

Using the decomposition (\ref{ADMlabel}) of $e^{a}_{t}$ along the $N$ and $N^{i}$ the action can
be rewritten as
\begin{equation}\label{Improved}
 \tilde{I}_{EH} = \int_{\mathcal{M}} \dot e^{a}_{i} \, \Omega^{\, i\  j}_{a\ bc} \,
 \omega^{bc}_{j}+ N H_{\perp}+ N^{i} H_{i} + \omega^{ab}_{t} J_{ab} + B
\end{equation}
where $H_{\perp}$ and $H_{i}$ are the projections of $P_{a}$ along the $\eta^{a}$ and $e^{a}_{i}$
respectively. One can see that $N$ and $N^{i}$ are Lagrange multipliers. In spite of the physical
interpretation of $H_{\perp}$ and $H_{i}$ is similar to the standard $H_{\mu}$ in metric formalism
(See for instance \cite{Regge:1974zd}) their formal expressions differ, recall that torsion has
not yet been eliminated.

$B$ in Eq.(\ref{Improved}) stands for the boundary term
\[ B = \int_{\mathbb{R} \times \partial \Sigma}  (N \eta^{a} + N^{l} e^{a}_{l})  \left(2\,\Omega_{a\ bc}^{\, i\
j}\,\omega^{bc}_{j} \varepsilon_{imn}\right)\,dt\wedge dx^{m}\wedge dx^{n}.
\]

\section{Transformation}

In this section is introduced a useful decomposition which isolates the conjugate momenta of the
12 $e_{i}^{a}$'s, contained in the 18 $\omega _{i}^{ab}$'s. In addition this gives rise to others
6 auxiliary fields $\lambda _{mn}$ and their 6 conjugate fields $\rho ^{mn}$ ($\lambda _{mn}$ and
$\rho ^{mn}$ are symmetric and $m,n =1,2,3$ \cite{Contreras:1999je}). Consider
\begin{equation}\label{DecomposingTheMomentum}
\omega _{k}^{ab}=\Theta _{k\ \ j}^{ab\ c}\ \pi _{c}^{j}+U_{k}^{ab\ \ mn}\ \lambda _{mn}
\label{trans1}
\end{equation}
Here $\Theta $ and $U$ are given by
\begin{equation}
\Theta _{i\ \ j}^{ab\ c}=\frac{1}{8\sqrt{g}}[e_{i}^{[a}\eta
^{b]}e_{j}^{c}-e_{i}^{[a}e_{j}^{b]}\eta ^{c}-2e_{j}^{[a}\eta ^{b]}e_{i}^{c}],
\end{equation}
\begin{equation}
U_{k}^{ab\ \ mn}=\frac{1}{2}\delta _{i}^{(m}\ \epsilon ^{n)kl}\ e_{k}^{a}\ e_{l}^{b},
\end{equation}
where the square brackets indicate antisymmetrization. In addition one can introduce
\begin{equation}
V_{ab\ \ mn}^{k}=\frac{1}{g}\ E_{a}^{r}\ E_{b}^{s}\epsilon _{rs(m}\ \delta _{n)}^{i}.
\end{equation}
These objects satisfy the following relations
\begin{eqnarray}
\Omega _{ab\ c}^{k\ \ i}\ \Theta _{k\ \ j}^{ab\ d}=\delta _{d}^{c}\delta _{j}^{i},\nonumber   \\
\Omega _{ab\ c}^{k\ \ j}\ U_{k}^{ab\ \ mn}=0, \label{orto4} \\
\Theta _{k\ \ j}^{ab\ c}\ V_{ab\ \ mn}^{k}=0,\nonumber  \\
U_{k}^{ab\ \ mn}\ V_{ab\ \ pq}^{k}=\delta _{(pq)}^{(mn)}\nonumber .
\end{eqnarray}

One can think of $\Theta $ and $\Omega $ as a collection of twelve vectors - labeled by the
indices $(_{i}^{a})$ and $(_{a}^{i})$ respectively-, in an 18-dimensional vector space with
components $(_{j}^{ab})$, and $(_{ab}^{j})$, respectively. Analogously, $U$ and $V$ correspond to
other six vectors. In this way the orthonormal relations (\ref{orto4}) defines the completeness
relation
\begin{equation}
\Theta^{ab \ e}_{i \ \ l} \ \Omega_{cd \ e}^{j \ \ l} + U^{ab \ \ mn}_{i} \ V_{cd \ \ mn}^{j} =
\delta^{[a b]}_{[c d]} \delta^i_j ,  \label{completeness}
\end{equation}
or equivalently
\[
\left(\Theta \, U\right) \left(
\begin{array}{c}
\Omega\\
V
\end{array}\right)
= Id_{18\times 18}
\]
in the 18 dimensional space.

\section{Hamiltonian expression and the role of torsion}
Using the decomposition (\ref{DecomposingTheMomentum}) the action becomes
\begin{equation}\label{ImprovedDecomposed}
\tilde{I}_{EH} = \int_{\mathcal{M}} (\dot e^{a}_{i} \pi_{a}^{i}+ N H_{\perp}+ N^{i} H_{i} +
\omega^{ab}_{t} J_{ab})\, dt\wedge d^{3}x + B
\end{equation}
Here $\pi_{a}^{i}$ is indeed the conjugate momentum of $e^{a}_{i}$. Furthermore
Eq.(\ref{ImprovedDecomposed}) is a genuine action principle with the correct Poison bracket being
$H_{\perp}$, $H_{i}$ and $J_{ab}$ constraints of first class provided the fixing of the $e^{a}$ at
the boundary \cite{Banados:1997hs}. For their expressions in terms of the fields see
\ref{ExplicitExpressionsofConstr}. Finally the Hamiltonian of this theory reads
\begin{equation}\label{HamiltonianFinalExpression}
H=-\int_{\Sigma} \dot e^{a}_{i} \pi_{a}^{i}+ N H_{\perp}+ N^{i} H_{i} + \omega^{ab}_{t}
J_{ab}\,d^{3}x - \hat{B},
\end{equation}
where
\begin{equation}\label{Boundaryterm}
 \hat{B} = \int_{\partial \Sigma} \left( N \eta^{a} \pi^{i}_{a} + N^{l} e^{a}_{l}
\pi^{i}_{a}\right)\varepsilon_{imn}
    dx^{m}\wedge dx^{n}.
\end{equation}
Recall that given that on shell the constraints must be satisfied the Hamiltonian
(\ref{HamiltonianFinalExpression}) reduces to the boundary term $H_{on\, shell}= -\hat{B}$. This
last observation will be essential to develop an expression for the energy in the next sections.

Before to proceed is worth to stress the role of torsion in this action principle
(\ref{ImprovedDecomposed}). In \cite{Aros:2003bi} was introduced the transformations of the fields
\begin{equation}
e_{\mu }^{a}\rightarrow e_{\mu }^{a}
\end{equation}
\begin{equation}
\omega _{\mu }^{ab}\rightarrow \omega _{\mu }^{ab}(e)+ K_{\mu }^{ab}
\end{equation}
which separates the spin connection into $K_{\mu }^{ab}$, the contorsion tensor, and $\omega _{\mu
}^{ab}(e)$ a torsion free connection. Since a torsion free connection is only a functional of the
vierbein this transformation can be visualized as a transformation of the Lagrangian
(\ref{EHactionimproved}) given by
\[ p\dot{q}- H \rightarrow (P+f(q))\dot{q}- H \rightarrow P\dot{q}- H', \]
where $H'= d \mathbb{F}/dt - H$ which readily can be identified as a canonical transformation.
Using this in \cite{Aros:2003bi} was proven that path order integrals of both metric and first
order formalism are equivalent provided the respective momenta have been integrated out.

It must stress that although the similar results one can expect of both formalism, in four
dimensional first order formalism an off-shell not vanishing torsion is essential to have a
Hamiltonian formulation, and thus four dimensional first order gravity represents a completely
different scenario compared with the four dimensional metric gravity.

\section{Geometry and coordinates at the boundary}

The boundary $\mathbb{R}\times \partial\Sigma$ must have a metric of the form
\begin{equation}\label{ADMBoundary}
    ds^{2} = -N^{2} dt^{2} + h_{mn}(V^{m} dt + d\sigma^{m})(V^{n} dt + d\sigma^{n}),
\end{equation}
where $\sigma^{m}$, with $m=2,3$, are the coordinates of slice at $t=const.$ of this boundary.
Since the boundary can be described as a surface $x^{\mu}(t,\sigma^{m})$ one can define a set of
(co-)vectors which give rise to metric (\ref{ADMBoundary}). This set reads
\begin{eqnarray}
  e^{a}_{t} &=& N \eta^{a} + V^{m} e^{a}_{m} \nonumber\\
  e^{a}_{m} &=& e^a_{m}, \label{ADMBoundarylabel}
\end{eqnarray}
where the projections are made by
\[ V^{m} = N^{i} \left.\frac{\partial \sigma^{m}}{\partial x^{i}}\right|_{\mathbb{R}\times
\partial\Sigma} \textrm{ and } e^{a}_{m} =  e^{a}_{i}\left.\frac{\partial x^{i}}{\partial
\sigma^{m}}\right|_{\mathbb{R}\times \partial\Sigma}
\]

To complete this analysis usually is introduced the unitarian vector $n^{a}$ which is normal to
the boundary $\mathbb{R}\times \partial \Sigma$, namely it satisfies
\[
n_{a}\eta^{a}=0 \textrm{ , } n_{a} e^{a}_{m} = 0 \textrm{ and } n^{a}n_{a}=1,
\]
and in this case can be defined as
\begin{equation}\label{VectorN}
n_{a} = \frac{1}{6 \sqrt{\gamma}}\varepsilon_{abcd}\, \eta^{b} e^{c}_{m}
e^{d}_{n}\varepsilon^{mn},
\end{equation}
where $\gamma=N^{2}h$ is determinant of the induced metric on $\mathbb{R}\times
\partial\Sigma$ that can be read in the line element (\ref{ADMBoundary}) and is only a functional
of $\eta_{a}$ and $e^{a}_{m}$.
\section{Energy Momentum}
Using the projections (\ref{ADMBoundarylabel}) the boundary term reads
\begin{equation}\label{projectedBoundaryTerm}
B = \int_{\mathbb{R} \times \partial \Sigma} \left( N \eta^{a}  + V^{m} e^{a}_{m}
\right)(\pi_{a}\cdot n)dt \wedge d^{2}\sigma.
\end{equation}
where $(\pi_{a}\cdot n)$ represents the projection $\pi_{a}^{i}n_{i}$ at the boundary.

Following the generalization of the Hamilton Jacobi equations proposed in \cite{Brown:1992br} one
can define an expression for the energy. Since the fields at the boundary are
$(N,V^{m},e^{a}_{m})$ here it is advisable to directly variate the action with respect each of
dynamical fields
\begin{eqnarray}
\frac{\delta \tilde{I}}{\delta N} &=&  \eta^{a}(\pi_{a}\cdot n)\label{variationAttheBoundary},\\
\frac{\delta \tilde{I}}{\delta V^{m}} &=&  e^{a}_{m} (\pi_{a}\cdot n)\nonumber,\\
\frac{\delta \tilde{I}}{\delta e^{a}_{m}} &=& \tau_{a}^{m}\nonumber
\end{eqnarray}
Note that $\tau^{m}_{a}$ is not a squared matrix, thus this definition is certainly different form
the metric one.

A definition of energy is given by integrating Eq.(\ref{variationAttheBoundary}), thus
\begin{equation}\label{Energy}
    E = -\int_{\partial \Sigma} \eta^{a}(\pi_{a}\cdot n)d^{2}\sigma,
\end{equation}
which is straightforward to show that it indeed recovers the mass for Schwarzschild or Reissner
N{\o}rdstrom solutions.

Likewise one can define the \textit{momentum}
\begin{equation}\label{Momentum}
    P_{m} = \int_{\partial \Sigma} e^{a}_{m} (\pi_{a}\cdot n) d^{2}\sigma,
\end{equation}
and an intrinsic \textit{energy momentum} tensor
\begin{equation}\label{IntrisicEnergyMomentumTensor}
T^{m}_{a} = \int_{\partial \Sigma}d^{2}\sigma \tau_{a}^{m}
\end{equation}

Following \cite{Brown:1992br} an energy density  can be defined as $e=-\eta^{a}(\pi_{a}\cdot n)$
as well as a momentum density $p_{m}=e^{a}_{m} (\pi_{a}\cdot n) $. Using these definitions the
Hamiltonian can be written as
\begin{equation}\label{HamiltonianInterpreted}
H = H_{\textrm{bulk}} + \int_{\partial \Sigma} ( e N - V^{m}p_{m})\, d^{2}\sigma,
\end{equation}
which is the first order version of the expression obtained in \cite{Brown:1992br}.

In the derivation within the metric formalism in \cite{Brown:1992br} the expressions in principle
were not only functional of the phase space variable, giving rise to an ambiguity, which was
removed by restricting expressions to satisfy that condition. Unlike this here the derivation was
made only after stating a truly Hamiltonian prescription for first order gravity, thus expressions
are only functional of the phase space variable by construction, and thus the ambiguity never
arose.

\section{Canonical ensemble action}
The variation of the action (\ref{ImprovedDecomposed}) can be cumbersome in terms of the phase
space fields, however recognizing that on shell the variation is merely given by
Eq.(\ref{VariationImproved}) one obtains that
\begin{equation}\label{FixofTheEnsemble}
\delta \tilde{I} = \int_{\Sigma_{\pm}} \pi^{i}_{a} \delta (e^{a}_{i}) + \int_{\mathbb{R}\times
\partial \Sigma} (e\delta N -  p_{m} \delta V^{m}  + \tau^{m}_{a} \delta e^{a}_{m})\,dt\wedge
d^{2}\sigma.
\end{equation}

The first term basically represents the generalization of the standard $p\,\delta
x|_{t_{i}}^{t_{f}}$ in any $0+1$ Lagrangian, in this case in the lids $\Sigma_{\pm}$.

The rest of Eq.(\ref{FixofTheEnsemble}) however demonstrates that in the variational principle
\eref{EHactionimprovedsplit} the energy, as defined in Eq.(\ref{Energy}), is not fixed, but the
lapse $N$. Furthermore the fixing of $N$ in turns fixes the scales of time, and thus the period in
the Euclidean version of the $\mathcal{M}$. This last result is equivalent to fix the temperature
which together with the unconstrained energy in Eq.(\ref{FixofTheEnsemble}) prove that action
(\ref{ImprovedDecomposed}) is indeed suitable for the canonical ensemble.

\section{Microcanonical boundary term}

To transform the action $\tilde{I}_{EH}$ into the microcanonical ensemble action is necessary to
add a boundary term that change the boundary conditions from fixing the period to fixing the
energy density $e$ of the system. This is simply achieved by subtracting from
Eq.(\ref{ImprovedDecomposed}) the term
\[\int_{\partial \Sigma} ( e N - V^{m}p_{m})\, d^{2}\sigma, \]
which is the boundary term $B$. This result leads to the new the action principle
\begin{equation}\label{Microcanonicalaction}
\hat{I}_{EH} = \int_{\mathcal{M}} (\dot e^{a}_{i} \pi_{a}^{i}+ N H_{\perp}+ N^{i} H_{i} +
\omega^{ab}_{t} J_{ab})\, dt\wedge d^{3}x,
\end{equation}
which is suitable for the microcanonical ensemble.

\section{Summary and prospects}

In this work a suitable action for the canonical ensemble in four dimensional first order gravity
has been found. Given this canonical ensemble action is direct to determine the corresponding
action for microcanonical ensemble as well.

Although the fundamental difference that the presence of torsion produces, the results reproduces
those of metric gravity. For instance the mass is recovered though part of the momentum, actually
the part the allows to compute the Dirac brackets, depends on torsion. This is in agreement with
every previous result, where metric gravity has been recovered in the aspect in discussion by
first order gravity on shell. The results in this work give even more reasons to study this
alternative theory of gravity.

One remarkable result in first order gravity is that once momenta $\pi^{i}_{a}$ are integrated out
the resulting expression is the same obtained in the metric formalism once the momenta, usually
denoted $\pi^{ij}$, are integrated out \cite{Aros:2003bi}. However both results -first order and
metric - were made ignoring the boundary terms. It will be very interesting to address the same
computation in first order gravity considering the presence of those boundary terms. Results in
the metric formalism considering the boundary terms are very promising and for instance are
connected the entropy of black holes \cite{Brown:1997rp}. One can expect that a similar result can
be achieved in first order gravity.

There are some aspects in first order gravity whose connection with the results in this work
should be fruitful to analyze. The introduction of a negative cosmological constant is interesting
since there are known results to compare with \cite{Aros:1999id}, in particular since the boundary
conditions in that work are radically different. With a negative cosmological constant the action
needs to be regularized in order to its Noether charge be finite, so in even dimensions, in
particular in four dimensions, was introduced the asymptotically locally AdS condition. The
boundary condition in \cite{Aros:1999id}, which is a mixed of Dirichlet and Neumann conditions, is
such that the boundary term of the variation vanish for arbitrary variation of the fields at the
spatial infinity. Remarkably that action is canonical ensemble action as well \cite{Aros:2001gz}.
The connection of that boundary condition in a Hamiltonian frame is an interesting direction to
continue the present work.

Another interesting issue to be address elsewhere is the role of boundary conditions on the
horizon in this Hamiltonian approach. Using a Noetherian approach those boundary conditions
determine the temperature in a completely different way as done here, and thus it would be very
interesting to explore the connection.

\appendix

\section{Explicit expressions}\label{ExplicitExpressionsofConstr}

The different constraints $H_{\perp }$, $H_{i}$, and $J_{ab}$ can be written explicitly as
\begin{equation}
\fl \frac{1}{2}H_{\perp }=\eta ^{a}\partial _{i}\pi _{a}^{i}-\frac{1}{2}%
E_{d}^{s}\partial _{[l}e_{s]}^{d}\eta ^{b}\pi _{b}^{l}-G_{\perp ij}^{ab}\pi _{a}^{i}\pi
_{b}^{j}-g^{3/2}G^{mnpq}\lambda _{mn}^{0}\lambda _{pq}^{0},
\end{equation}

\begin{eqnarray}
 \fl N^{m}H_{m}= N^{m}\,\left[\frac{1}{2}\left(g^{-1}E_{d}^{s}\partial
_{i}e_{k}^{d}\varepsilon_{mls}\varepsilon ^{ijk}e_{j}^{b}-E_{d}^{s}\partial
_{[m}e_{s]}^{d}e_{l}^{b}+\eta _{d}\partial _{[m}e_{l]}^{d}\eta ^{b}\right)\pi
_{b}^{l}\right.\nonumber \\
 \lo + \left. e_{m}^{a}\partial _{i}\pi _{a}^{i}+G_{mij}^{ab}\pi _{a}^{i}\pi _{b}^{j}+%
\frac{1}{2}N^{(m}e_{i}^{a}e_{j}^{b}J_{ab}\epsilon ^{ijn)}\lambda _{mn}^{0} \right]\\
\lo  -N^{i}\omega _{i}^{ab}\ J_{ab}\nonumber,
\end{eqnarray}
and
\begin{equation}
J_{ab}=2\varepsilon _{abcd}\frac{\partial e_{j}^{c}}{\partial x^{i}}%
e_{k}^{d}\varepsilon ^{ijk}-\frac{1}{2}(\pi _{a}^{i}e_{bi}-\pi _{b}^{i}e_{ai}),
\end{equation}
where
\begin{equation}
\lambda _{pq}^{0}=\frac{1}{2g}G_{pqmn}E_{a}^{(m}\partial _{i}e_{j}^{a}\epsilon ^{ijn)},
\label{lambda0}
\end{equation}
\begin{equation}
G_{\perp ij}^{ab}=\frac{1}{16\sqrt{g}}[%
e_{i}^{a}e_{j}^{b}-2e_{j}^{a}e_{i}^{b}-g_{ij}\eta ^{a}\eta ^{b}],
\end{equation}
and
\begin{equation}
G_{mij}^{ab}=\frac{1}{16\sqrt{g}}[g_{ij}\eta ^{a}e_{m}^{b}+2g_{im}(e_{j}^{a}\eta
^{b}-e_{j}^{b}\eta ^{a})].
\end{equation}

\ack This work is partially funded by grants FONDECYT 1040202 and DI 06-04. (UNAB). Rodrigo Aros
would like to thank Abdus Salam International Centre for Theoritical Physics for his associate
award.

\vspace{0.3in}

%\bibliographystyle{jhep}
%\bibliography{myXbib}
\providecommand{\href}[2]{#2}\begingroup\raggedright\endgroup

\end{document}